\def\year{2019}\relax
\documentclass[letterpaper,anonymous=false]{article} 
\usepackage{aaai19}  
\usepackage{times}  
\usepackage{helvet} 
\usepackage{courier}  
\usepackage[hyphens]{url}  
\newcommand{\algname}[1] {{\fontfamily{cmtt}\selectfont {#1}}}
\usepackage{amsmath,cite,url}
\usepackage{graphicx}
\usepackage{color}
\usepackage{todonotes}

\usepackage{graphicx}  
\usepackage{booktabs} 
\usepackage{caption}
\usepackage{subcaption}
\usepackage{enumerate}
\usepackage{multirow}
\usepackage{natbib}

\usepackage{graphicx} 
\usepackage{graphicx}  
\title{Investigating Potential Factors Associated with Gender Discrimination in Collaborative Recommender Systems}
\author{
Masoud Mansoury \thanks{This author also has affiliation in School of Computing, DePaul University, Chicago, USA, mmansou4@depaul.edu.} \\
{Eindhoven University of Technology}\\
m.mansoury@tue.nl
\newline \\
\newline \\
\textbf{\large Arman Dehpanah}\\
{DePaul University}\\
adehpanah@depaul.edu
\And 
Himan Abdollahpouri\\
{University of Colorado Boulder}\\
himan.abdollahpouri@colorado.edu
\newline \\
\newline \\
\textbf{\large Mykola Pechenizkiy}\\
{Eindhoven University of Technology}\\
m.pechenizkiy@tue.nl
\And
Jessie Smith\\
{University of Colorado Boulder}\\
jessie.smith-1@colorado.edu
\newline \\
\newline \\
\textbf{\large Bamshad Mobasher}\\
{DePaul University}\\
mobasher@cs.depaul.edu
}

\begin{document}

\maketitle

\begin{abstract}

The proliferation of personalized recommendation technologies has raised concerns about discrepancies in their recommendation performance across different genders, age groups, and racial or ethnic populations. This varying degree of performance could impact users' trust in the system and may pose legal and ethical issues in domains where fairness and equity are critical concerns, like job recommendation. In this paper, we investigate several potential factors that could be associated with discriminatory performance of a recommendation algorithm for women versus men. We specifically study several characteristics of user profiles and analyze their possible associations with disparate behavior of the system towards different genders. These characteristics include the anomaly in rating behavior, the entropy of users’ profiles, and the users’ profile size. Our experimental results on a public dataset using four recommendation algorithms show that, based on all the three mentioned factors, women get less accurate recommendations than men indicating an unfair nature of recommendation algorithms across genders. 
\end{abstract}
\section{Introduction}

Recommender systems are powerful tools for predicting users' preferences and generating personalized recommendations. It has been shown that these systems, while effective, can suffer from lack of fairness in their recommendation output. The generated recommendations by these systems are, in some cases, biased against certain groups (\cite{ekstrand2018}). This discrimination among users could negatively affect users' satisfaction, and at worst, can lead to or perpetuate undesirable social dynamics. 

Unfair recommendation is often defined as inconsistent performance of a recommendation algorithm for different groups of users. For example, suppose we have two user groups $A$ and $B$. The recommender system would be considered unfair if it delivered significantly and consistently better recommendations for group $A$ than $B$ or vise versa. This unfairness can be even more problematic if it causes unintentional discrimination against individuals in protected classes such as gender, race, or age \footnote{https://www.eeoc.gov/laws/types/}. In this paper, we focus on one protected class: gender.


\cite{mansoury2019relationship} argue that the anomaly of users' rating behavior could be one of the reasons for receiving unfair recommendations. In this work, the authors show that groups with higher anomaly in their rating behavior receive less calibrated and, hence, less fair recommendations. However, they do not test their claim on protected groups, such as men versus women. In this paper, we explore this idea further.
 
 We first show that the degree of anomalous rating behavior in male and female's profile does not explain why these two groups receive different levels of accuracy in their recommendations. As we show in the experimental results section, female groups still receive significantly less accurate recommendations than males with the same level of anomalous rating behavior. We then explore two other factors that could potentially be associated with poor recommender system performance for women. 
 We will describe each of these factors in detail in the following sections.

\section{Related work}
Previous research has raised concerns about discrepancies in recommendation accuracy across different genders (\cite{yao2017,zhu2018,kamishima2011}). For instance, \cite{ekstrand2018} show that women on average receive less accurate, and consequently, less fair recommendations than men using a movie dataset.

It is crucial to figure out which factors might be leading to unfairness in recommendations to inform potential solutions to this problem. \cite{abdollahpourirmse1,abdollahpouriWSDM} show that popularity bias can negatively affect the performance of the recommendations. In that work, authors show that algorithms with higher popularity bias give less accurate recommendations to the users. 

In a more recent work,  \cite{mansoury2019relationship} argued that anomalous rating behavior could be one of the reasons for unfair recommendations. In their experiments, there is indeed a correlation between the rating anomaly and the performance of the recommendations for different user groups. However, the authors did not specify user groups based on sensitive attributes (e.g., men versus women), but rather a general grouping based on the degree of anomaly of user profiles. In our work, given gender as a sensitive attribute, we show that anomalous rating behavior does correlate with recommendation performance for men. However, as we see in the experimental results section, it does not explain why women, on average, receive less accurate recommendations.


\begin{figure*}[htp]
  \centering
  \begin{subfigure}[b]{0.99\textwidth}
        \includegraphics[width=0.24\textwidth]{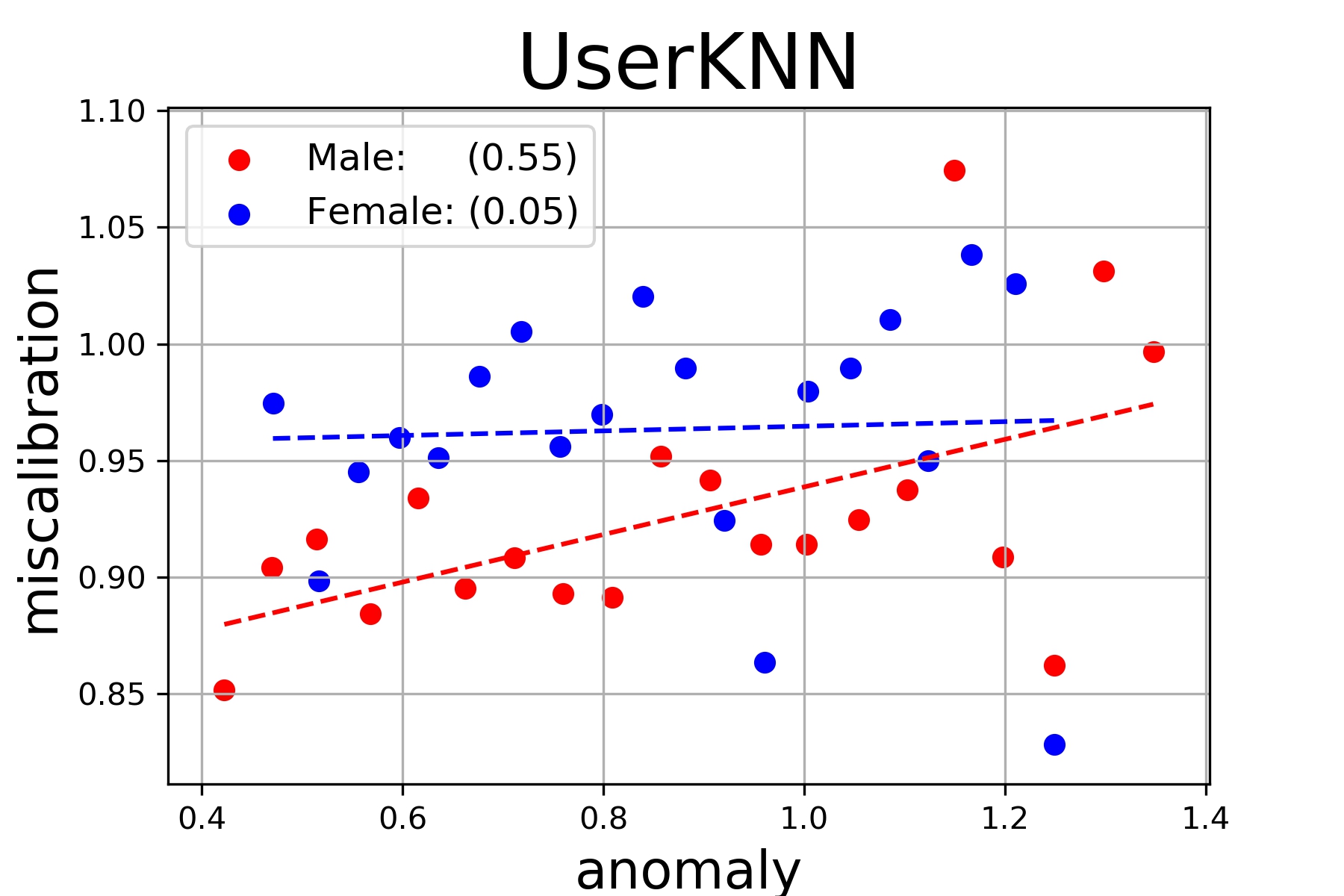}
        \includegraphics[width=0.24\textwidth]{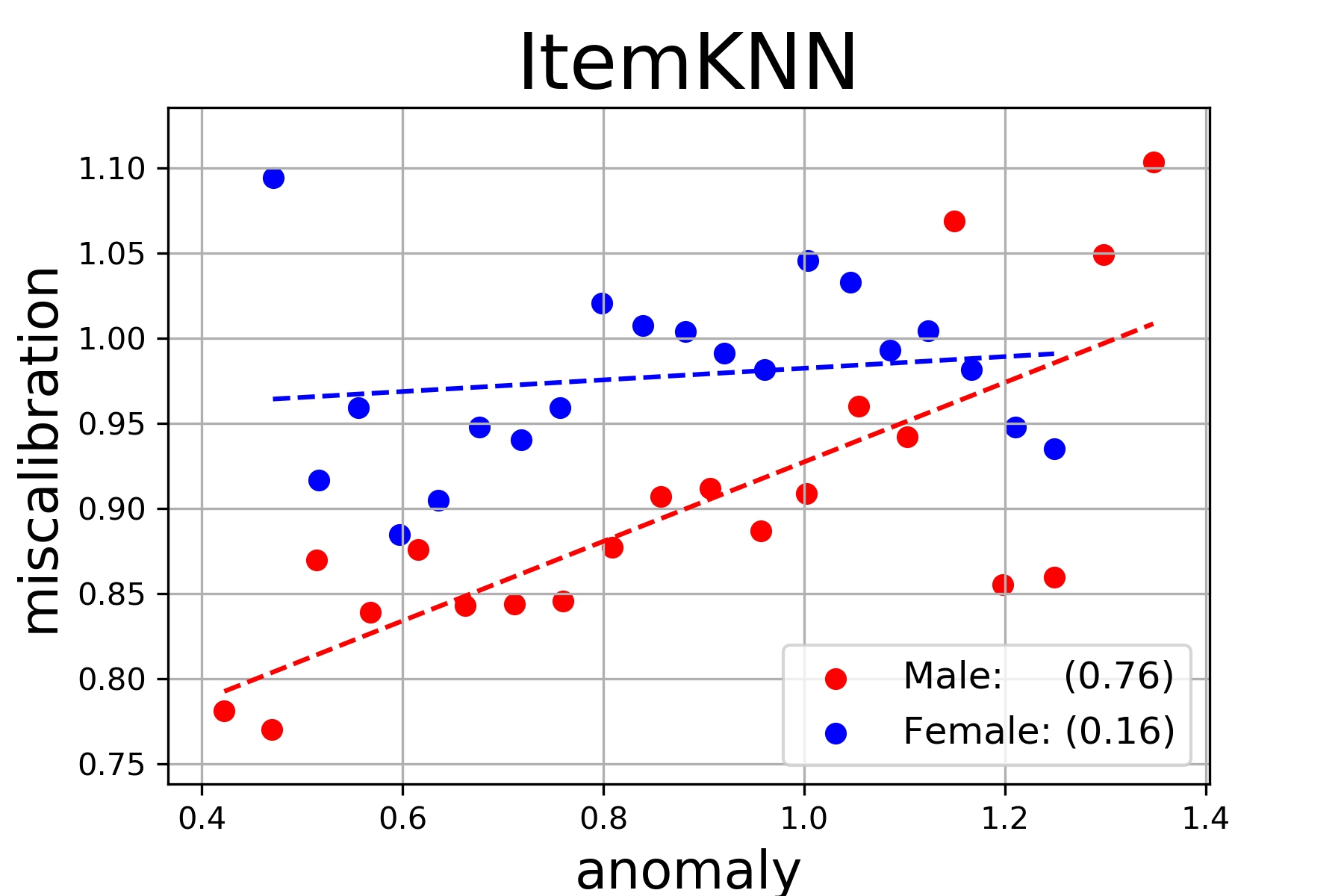}
        \includegraphics[width=0.24\textwidth]{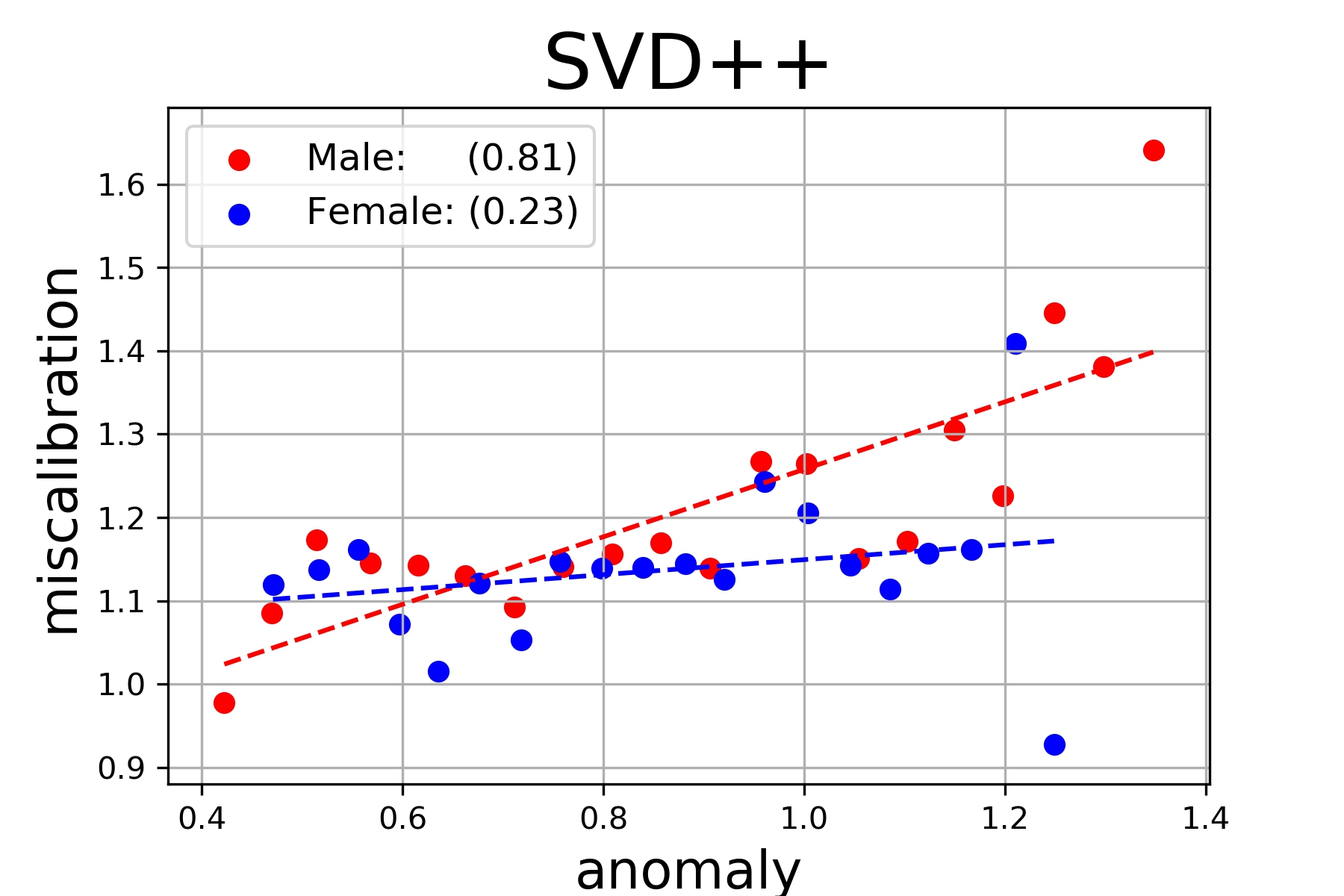}
        \includegraphics[width=0.24\textwidth]{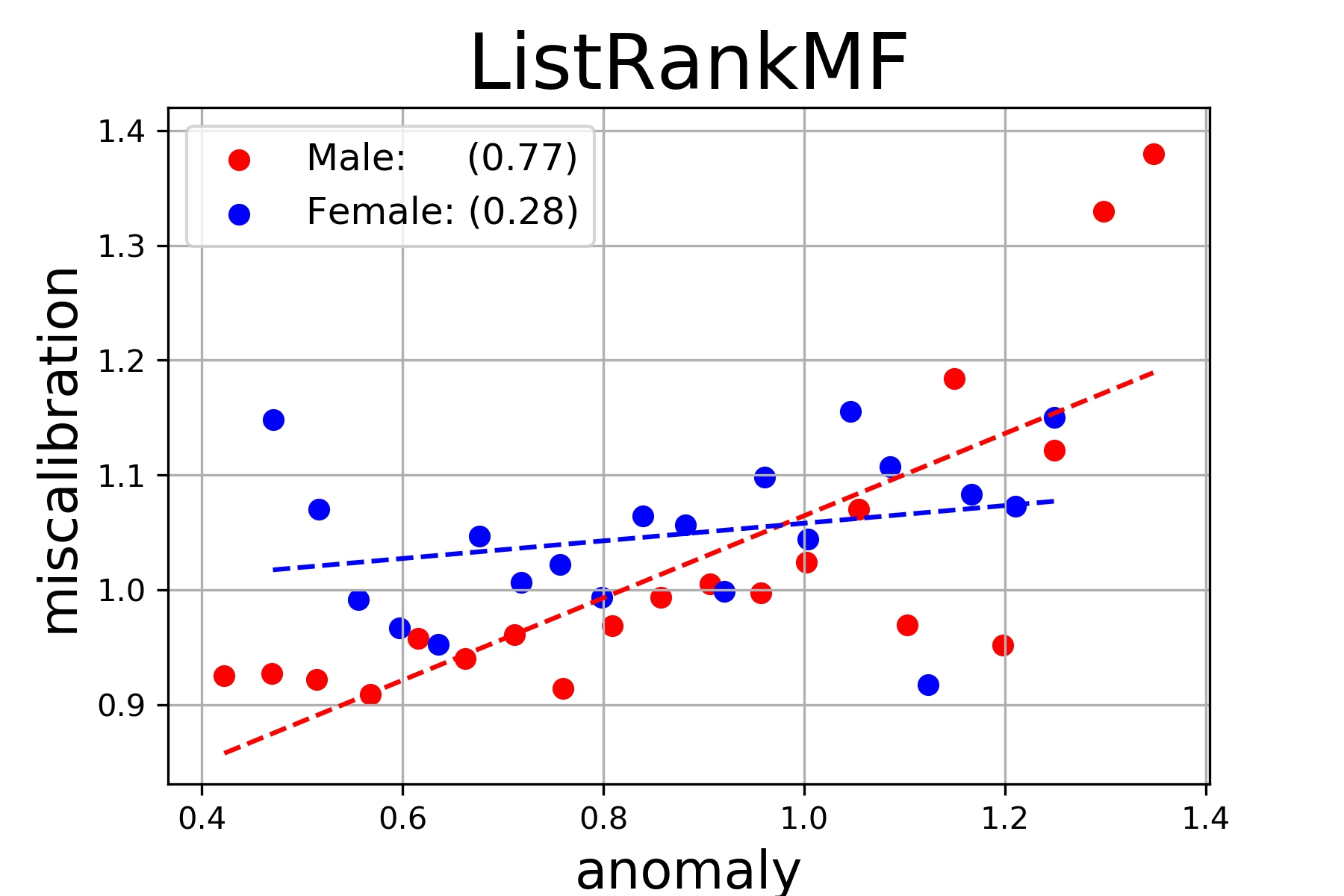}
        
  \end{subfigure}
  \begin{subfigure}[b]{0.99\textwidth}
        \includegraphics[width=0.24\textwidth]{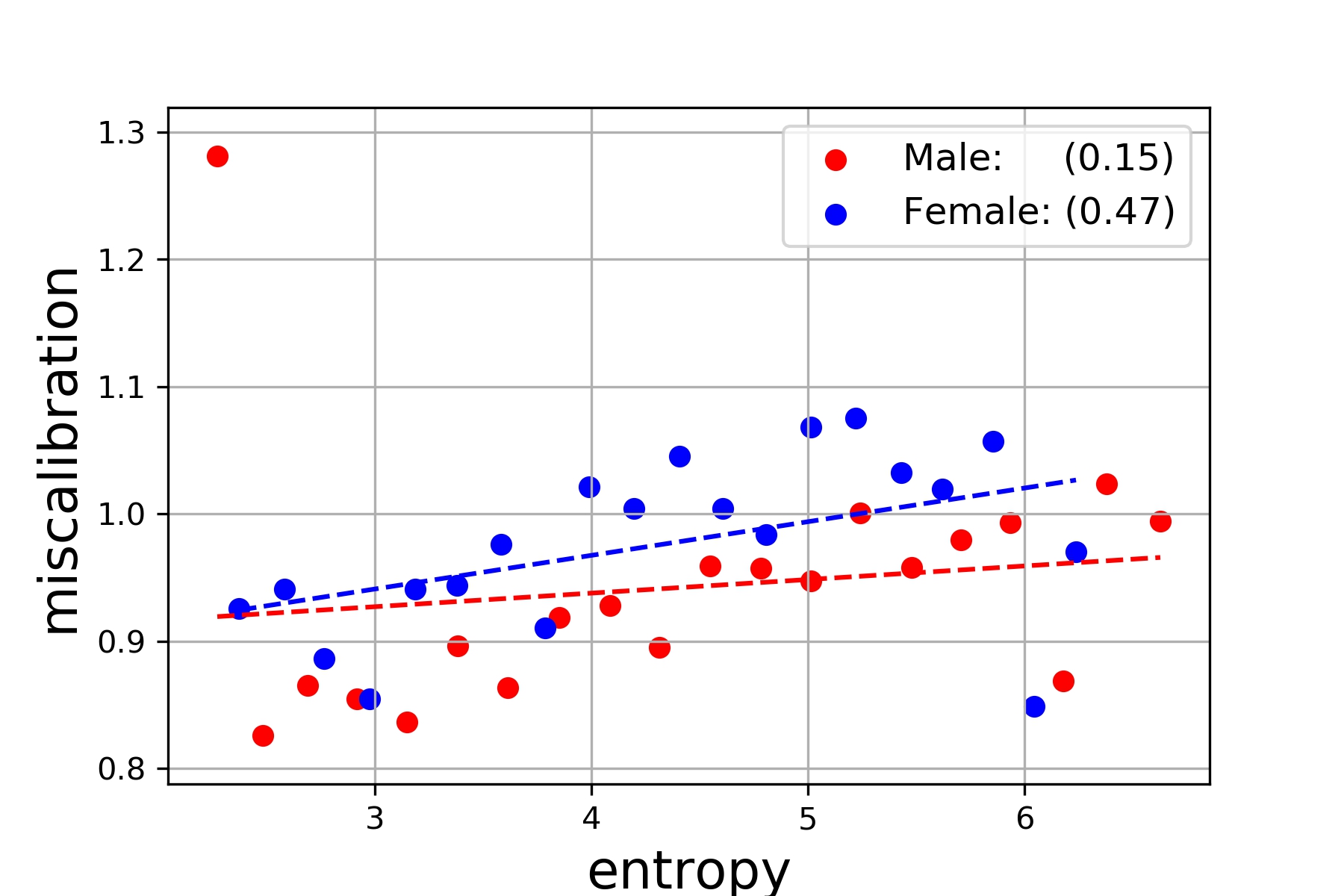}
        \includegraphics[width=0.24\textwidth]{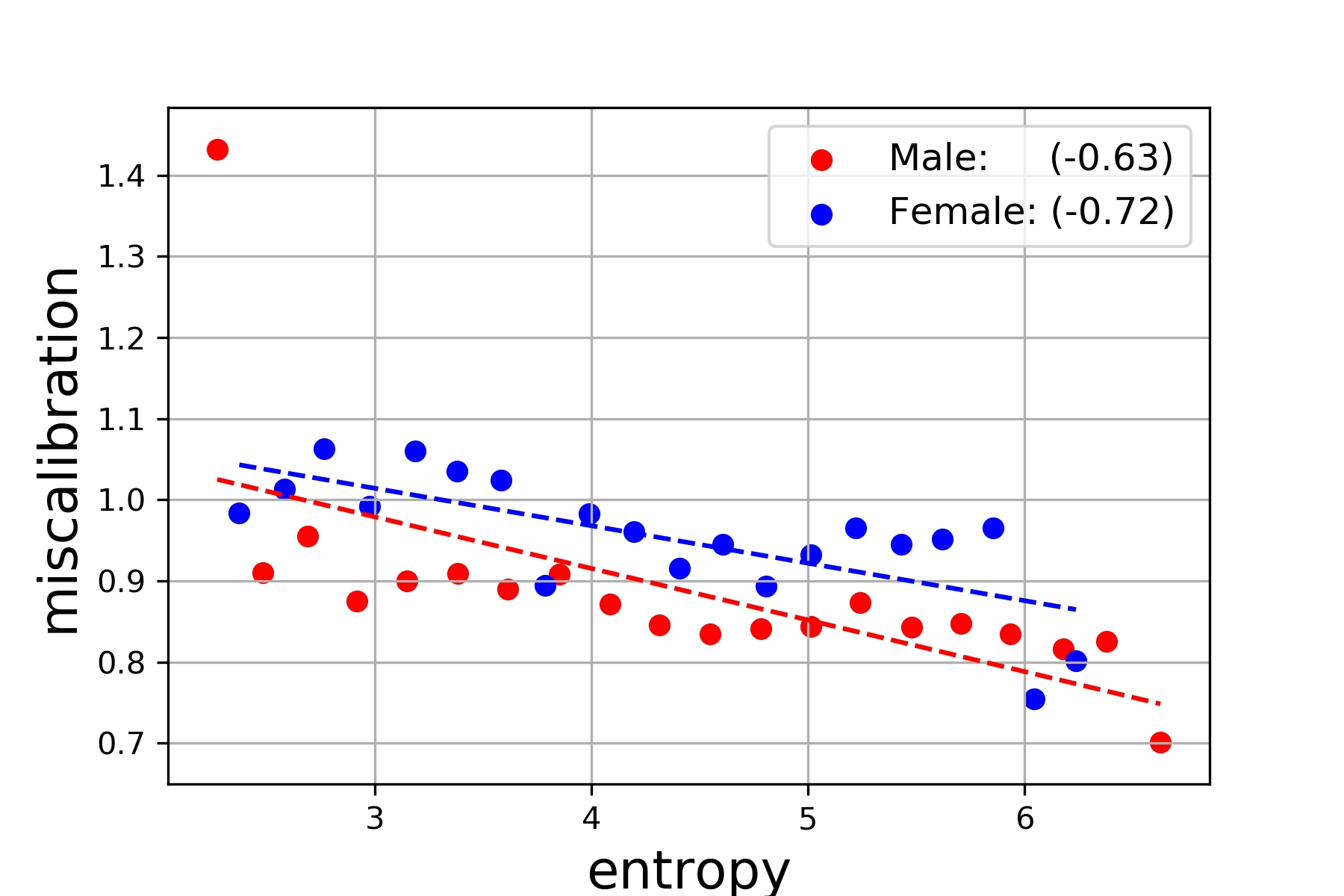}
        \includegraphics[width=0.24\textwidth]{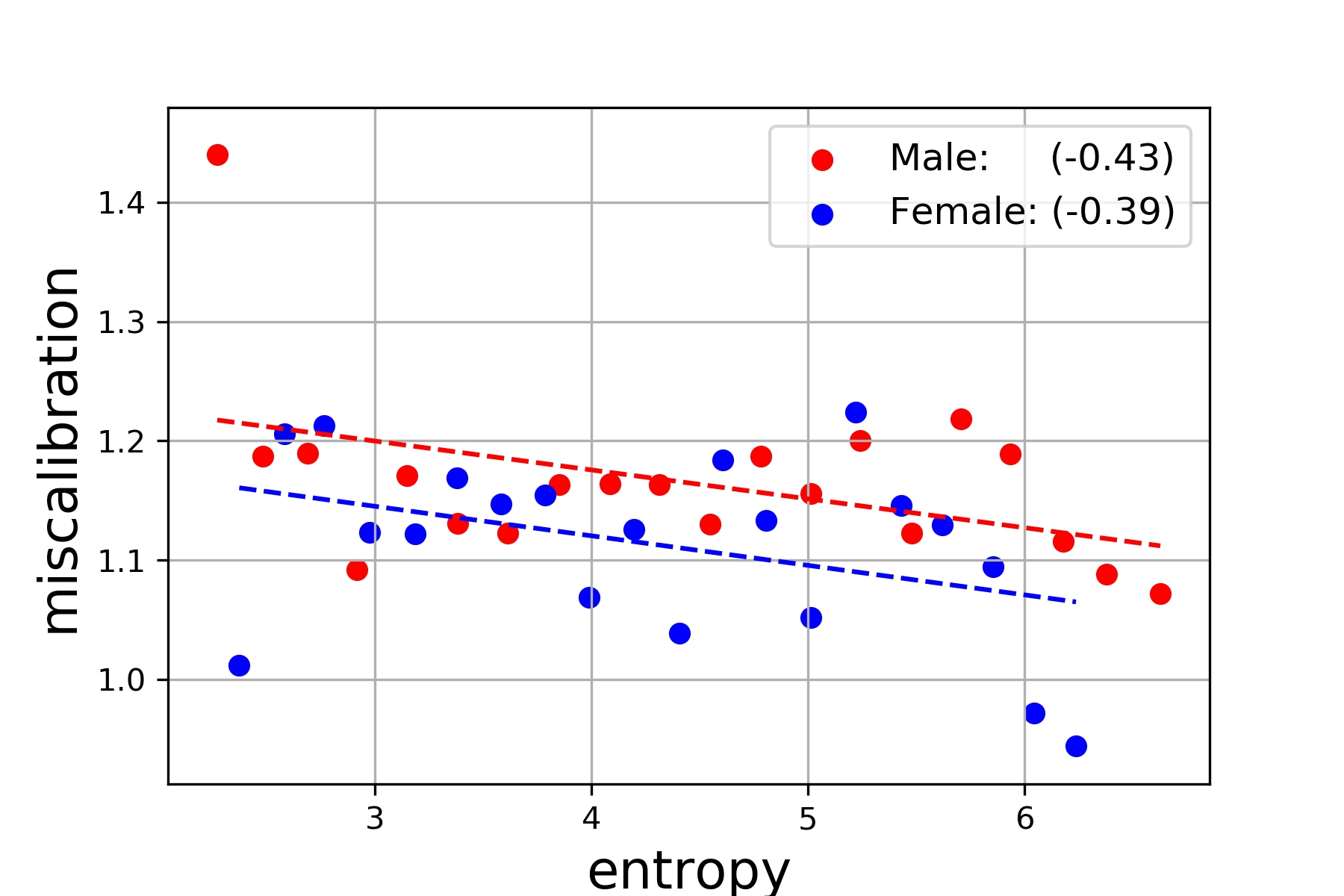}
        \includegraphics[width=0.24\textwidth]{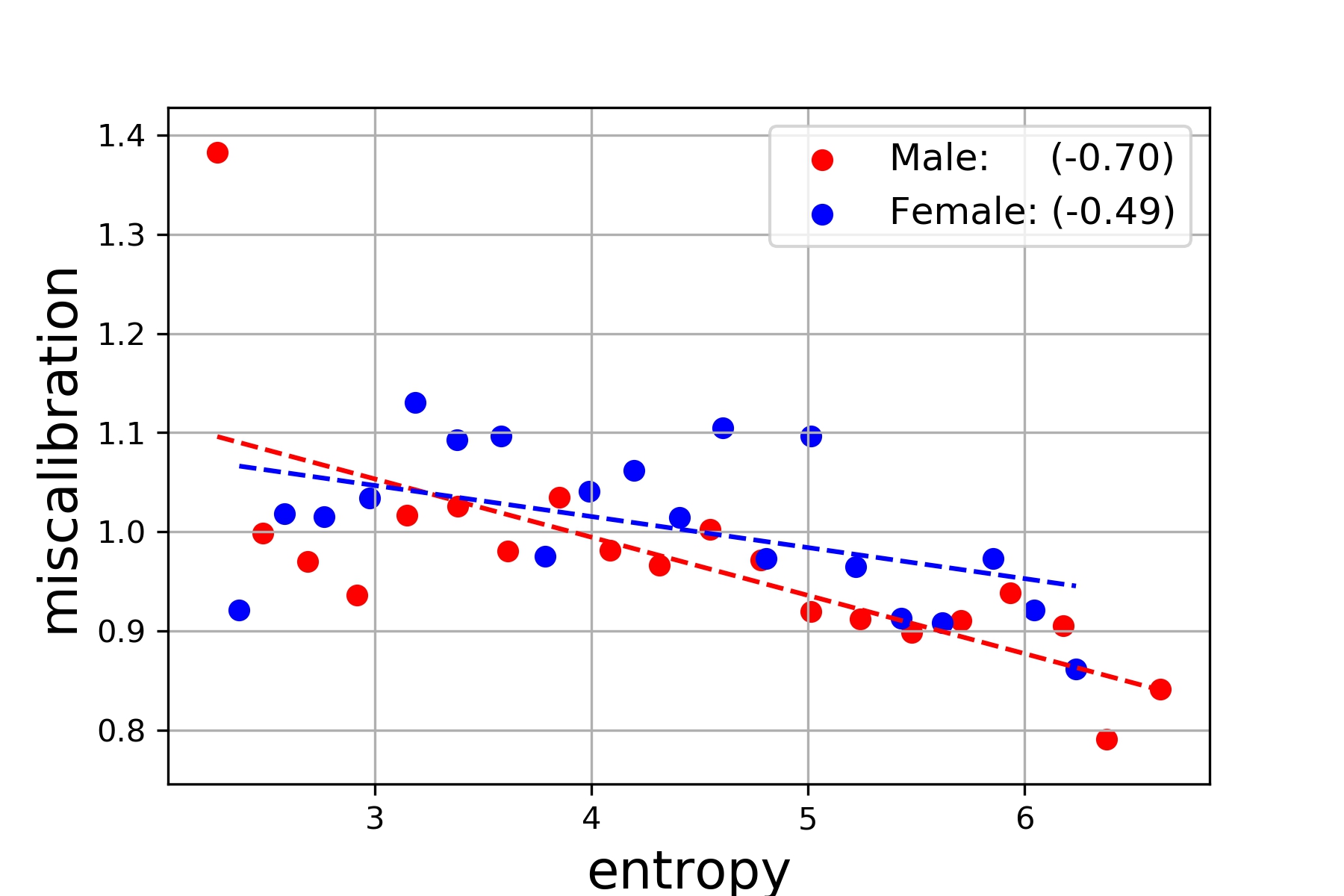}
  \end{subfigure}
  \begin{subfigure}[b]{1\textwidth}
        \includegraphics[width=0.24\textwidth]{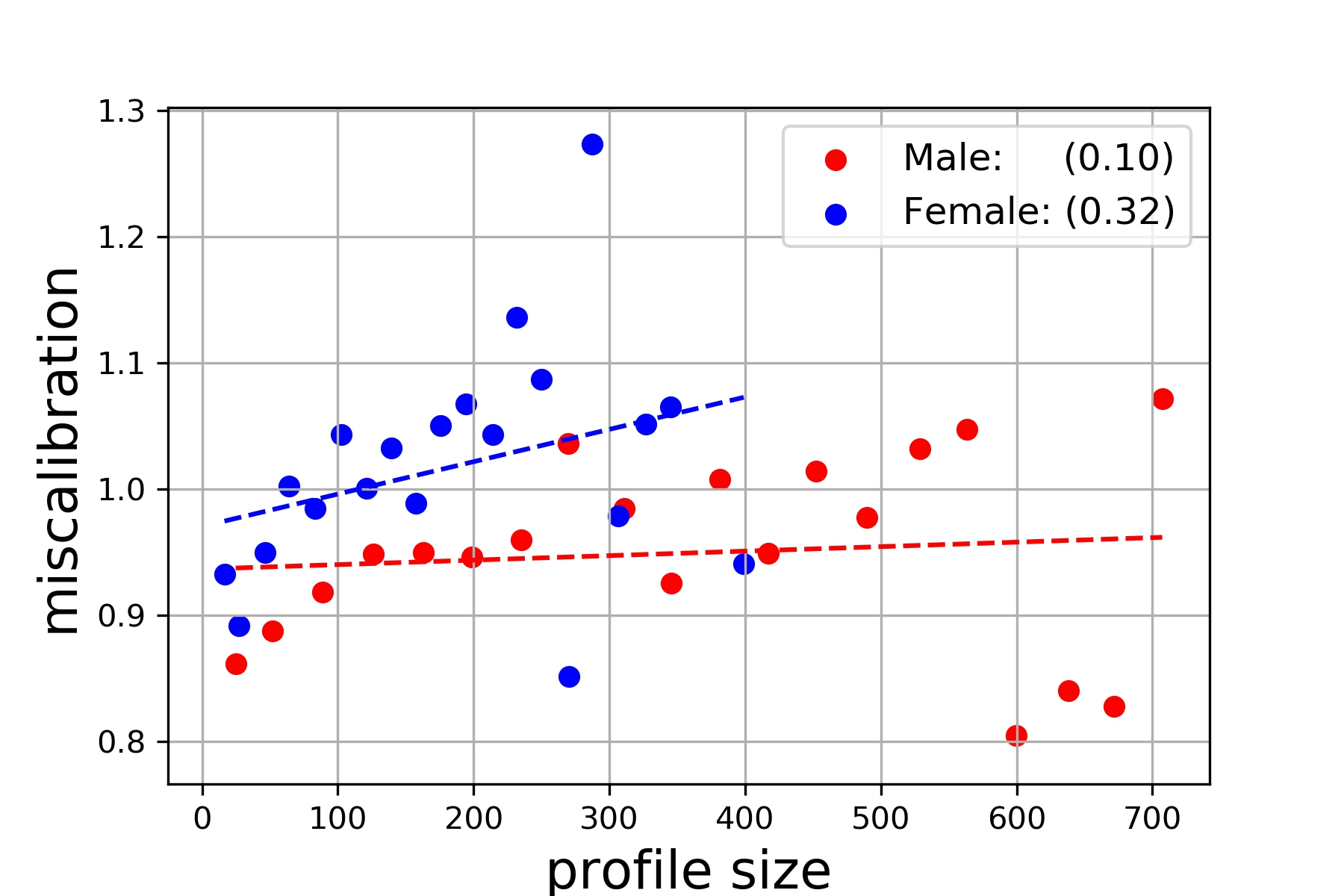}
        \includegraphics[width=0.24\textwidth]{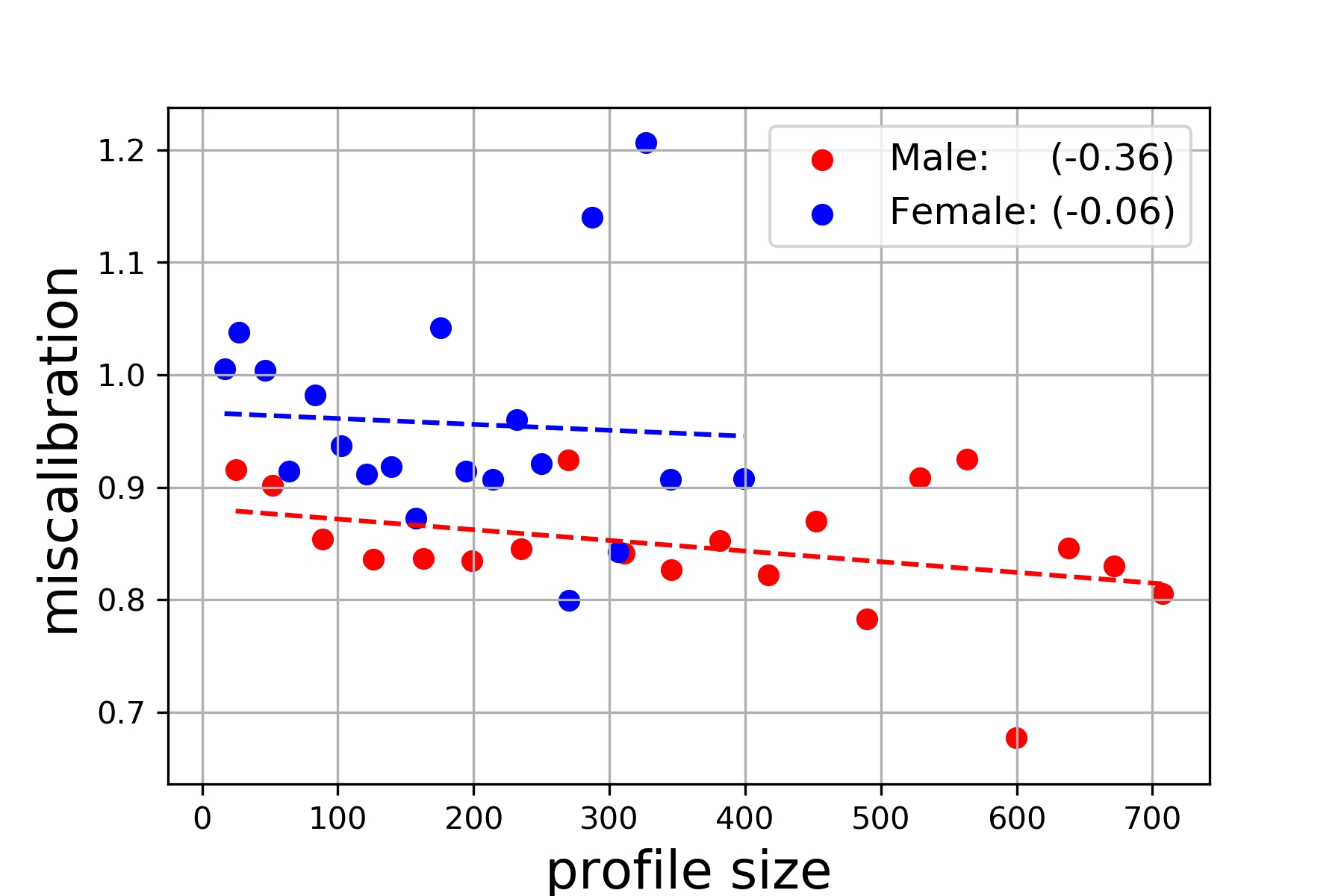}
        \includegraphics[width=0.24\textwidth]{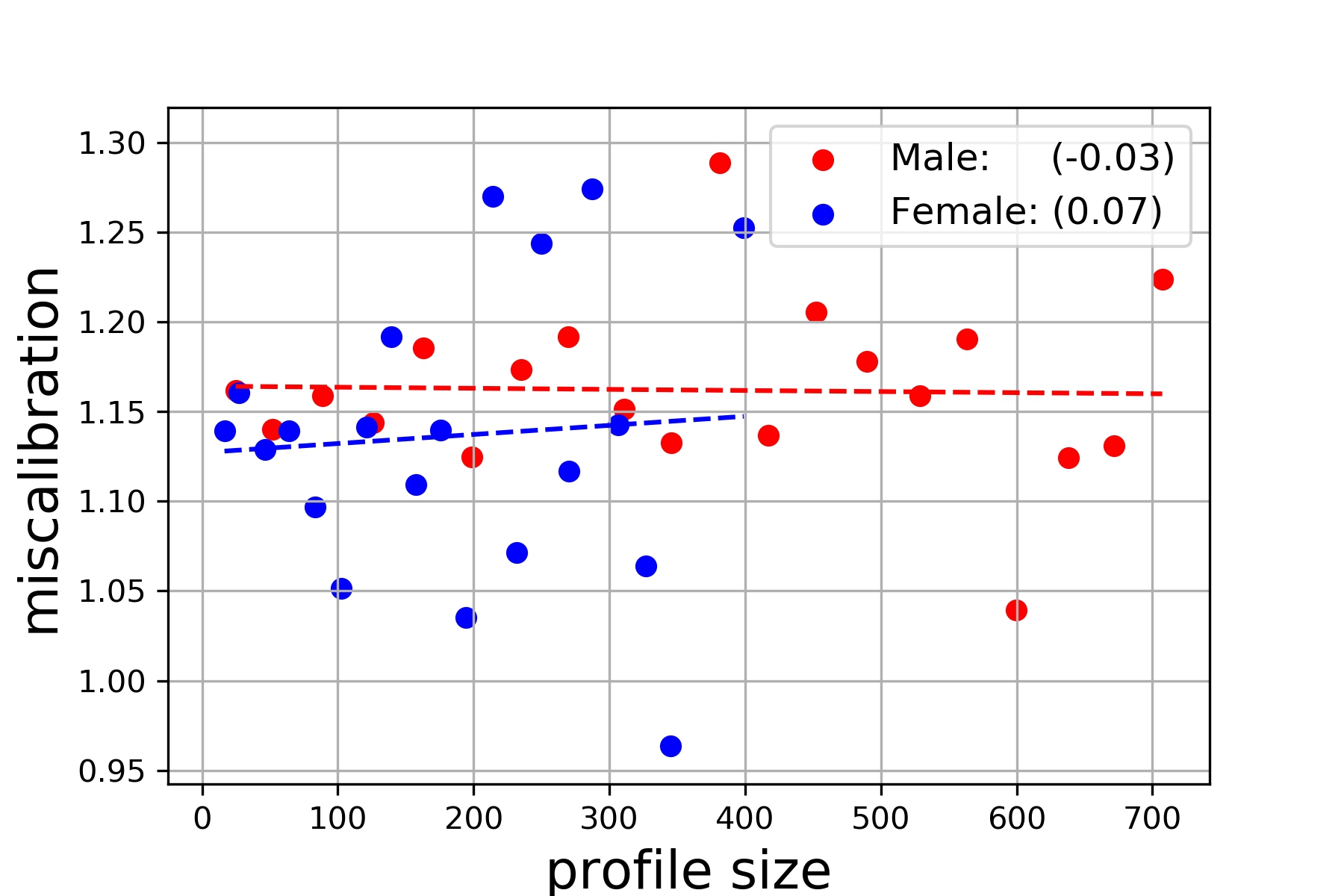}
        \includegraphics[width=0.24\textwidth]{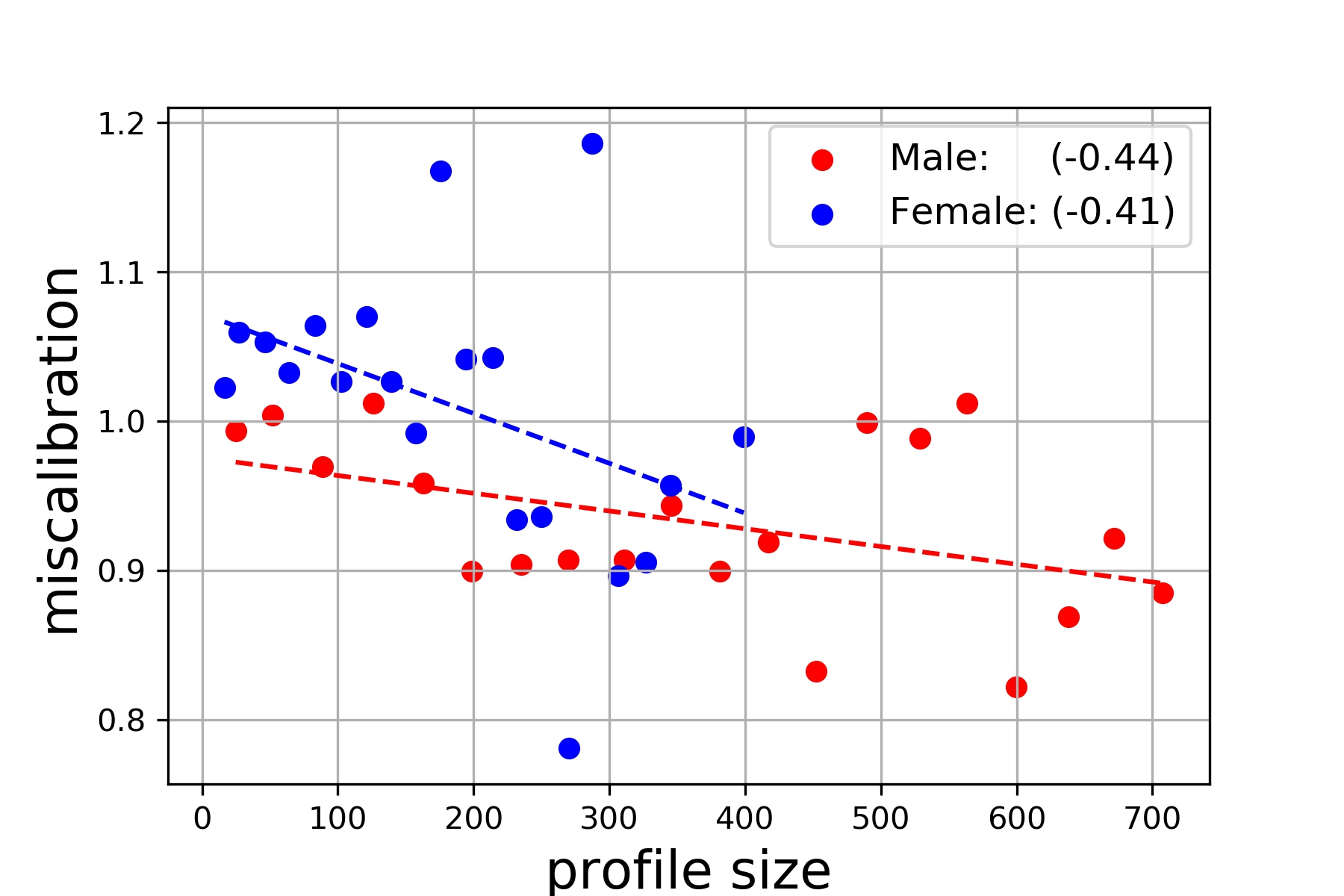}
  \end{subfigure}
\caption{The Correlation between anomaly, entropy, and size of the users' profiles and  miscalibration of the recommendations generated for them. Numbers next to the legends in the plots show the correlation coefficient for each user group.} \label{fig:corr_cons}
\end{figure*}

%
%
 %

\section{Factors associated with unfair recommendations}

In this section, we discuss three different factors that might lead to a poor recommendation performance. 

\begin{itemize}
    \item \textbf{Profile anomaly ($\mathcal{A}$):} As discussed in \cite{mansoury2019relationship}, one factor that could impact recommendation performance is the degree of anomalous rating behavior relative to other users. 
    The authors in this paper showed that users whose rating behavior is more consistent with other users in the system as a whole receive better recommendations than those who have more anomalous ratings. This happens because users who rate more in line with typical users are likely to find more matching items or users. We measure the degree of profile anomaly based on how similarly a user rates items compared to the majority of other users who have rated that item. Since collaborative filtering approaches use opinions of other users (e.g. similar users) for generating recommendations for a target user, it is highly possible that users with anomalous ratings receive less accurate recommendations. Given a target user, $u$, and $I_{u}$ as all items rated by $u$, profile anomaly of $u$ can be calculated as:


\begin{equation} \label{eq:consistency}
\begin{aligned}
\mathcal{A}_{u}=\frac{\sum_{i \in I_{u}}|r_{u,i}-\overline{r_{i}}|}{N_{u}}
\end{aligned}
\end{equation}

where $r_{u,i}$ is the rating given by $u$ to item $i$, $\overline{r_{i}}$ is  the average rating assigned to item $i$, and $N_{u}$ is the number of items rated by $u$ (i.e. the profile size of $u$).


    \item \textbf{Profile entropy ($\mathcal{E}$):}
    Another possible factor that could impact recommendation performance is how informative a user's profile is. The more diverse a user's ratings are, the higher their entropy is. For example, has the user only given high (or low) ratings to different items? Or are there a wide range of different ratings given by the user? We measure the entropy of user $u$'s profile as follows:
    
    \begin{equation}
        \mathcal{E}_{u}=-\sum_{v \in V}D_{u}(v)\log D_{u}(v)
    \end{equation}
    
    where $V$ is the set of discrete rating values (for example, 1,2,3,4,5) and $D_{u}$ is the observed probability distribution over those values in $u$'s profile. 
    
    \item \textbf{Profile size ($\mathcal{S}$):} The last factor we investigate in this paper is the profile size of each user. We believe users who are more active in the system (and have rated a larger number of items) receive better recommendations compared to those with shorter profiles. 
\end{itemize}

\section{Methodology}
For our experiments, we use the well-known MovieLens 1M (ML1M) dataset. In this dataset, 6,040 users provided 1,000,209 ratings (602,881 given by males and 197,286 given by females) on 3,706 movies. 
Table \ref{tab:stat} shows the specification of ML1M dataset for male and female users. As shown in this table, there are more male users in the dataset than female users. Moreover, on average, male users have larger profiles, and their profile entropy is also higher than female users. In addition, the average anomaly of male users' profiles is slightly lower than female users. 
\captionsetup[table]{skip=4pt}
\begin{table}[t!]
\small
\centering
\captionof{table}{Specification of ML1M for male and female users} \label{tab:stat}
\begin{tabular}{l|cccc}
\toprule
  & $\#users$ & $\overline{\mathcal{A}}$ & $\overline{\mathcal{E}}$ & $\overline{\mathcal{S}}$ \\
 \midrule
 Male & 4,331 & 0.781 & 4.174 & 139.2 \\
 Female & 1,709 & 0.808 & 3.995 & 115.4 \\
 \bottomrule
\end{tabular}
\end{table}

 We divide the dataset into training and test sets in an 80\% - 20\% ratio, respectively. The training set is then used to build the model. After training different recommendation algorithms, we generate recommendation lists of size 10 for each user in the test set.

We create 20 user groups separately for males and females by measuring different factors: degree of anomaly, entropy, and profile size, discussed further in the previous section. Specifically, we sort users based on each factor and then split them into 20 buckets in an ascending order. Users that fall within each bucket represent one group. In order to calculate the anomaly, entropy, profile size, precision, and miscalibration for each group, we average the corresponding measure over all the users in the group.


We run our experiments using four recommendation algorithms: user-based collaborative filtering (\algname{UserKNN}), item-based collaborative filtering (\algname{ItemKNN}), singular value decomposition (\algname{SVD++}), and list-wise matrix factorization (\algname{ListRankMF}).
All recommendation models are optimized using Grid Search over hyperparameters and the configuration with the highest precision is selected. The precision values for \algname{UserKNN}, \algname{ItemKNN}, \algname{SVD++}, and \algname{ListRankMF} are 0.214, 0.223, 0.122, and 0.148, respectively.
We used \textit{librec-auto} and LibRec 2.0 for all experiments (\cite{mansoury2018automating,Guo2015}).

\captionsetup[table]{skip=4pt}
\begin{table}[t!]
\small
\centering
\captionof{table}{Precision and miscalibration of recommendation algorithms for male and female users} \label{tab:accuracy}
\begin{tabular}{llllll}
\toprule
 \multirow{2}{*}{algorithm} & \multicolumn{2}{c}{Precision} & & \multicolumn{2}{c}{Miscalibration} \\ \cline{2-3}\cline{5-6}
 & Male & Female & & Male & Female \\
 \bottomrule
 \algname{UserkNN} & 0.235 & 0.162 & & 0.915 & 0.971 \\
 \algname{ItemkNN}  &  0.242 & 0.175 & & 0.874 & 0.973 \\
 \algname{SVD++} &  0.133 & 0.095 & & 1.156 & 1.130 \\
 \algname{ListRankMF} & 0.160 & 0.118 & & 0.970 & 1.032 \\
 \bottomrule
\end{tabular}
\end{table}

\section{Experimental Results}

Table \ref{tab:accuracy} shows the performance of recommendation algorithms for male and female users. In terms of precision, male users consistently receive more accurate recommendations than females and in terms of miscalibration, except for \algname{SVD++}, male users receive less miscalibrated (i.e. more calibrated) recommendations than females. Lower miscalibration for females than males on \algname{SVD++} shows an interesting result in our experiments that needs further investigation.

Figure~\ref{fig:corr_cons} shows the relationship between the degree of anomaly, entropy, and profile size for 20 user groups for both male and female users and the miscalibration of the recommendations they received. As we can see in the first row (anomaly vs miscalibration), in all algorithms except for \algname{SVD++}, the recommendations given to the female users have higher miscalibration (they are less calibrated) regardless of the anomaly of their ratings compared to the male user groups. Also, we can see that the positive correlation between profile anomaly and recommendation miscalibration discussed in \cite{mansoury2019relationship} can only be seen on male users. 
The second row of Figure~\ref{fig:corr_cons} shows the relationship between the entropy of the ratings and the miscalibration of their recommendations. Again, it can be seen that except for \algname{SVD++}, for all other algorithms, female user groups have higher miscalibration in their recommendations regardless of the amount of entropy of their ratings. 

Finally, the last row of Figure~\ref{fig:corr_cons} shows the correlation between the average profile size of different user groups and the miscalibration of their recommendations. Looking at this plot, we can see that there is no significant correlation between these two indicating the profile size of the users does not affect the miscalibration of their recommendations. However, except for \algname{SVD++}, again all algorithms have higher miscalibration for female user groups regardless of their profile size. It seems that \algname{SVD++} is indeed the \textit{fairest} algorithm among the four as it gives a comparable performance for both male and female users. It can also be seen that there is no data point for female groups when the value of the $x$ axis is larger than 400, meaning the largest average profile size for female groups is 400 while there are some male user groups with an average profile size of around 700.

\begin{figure*}[htp]
  \centering
  \begin{subfigure}[b]{0.99\textwidth}
        \includegraphics[width=0.24\textwidth]{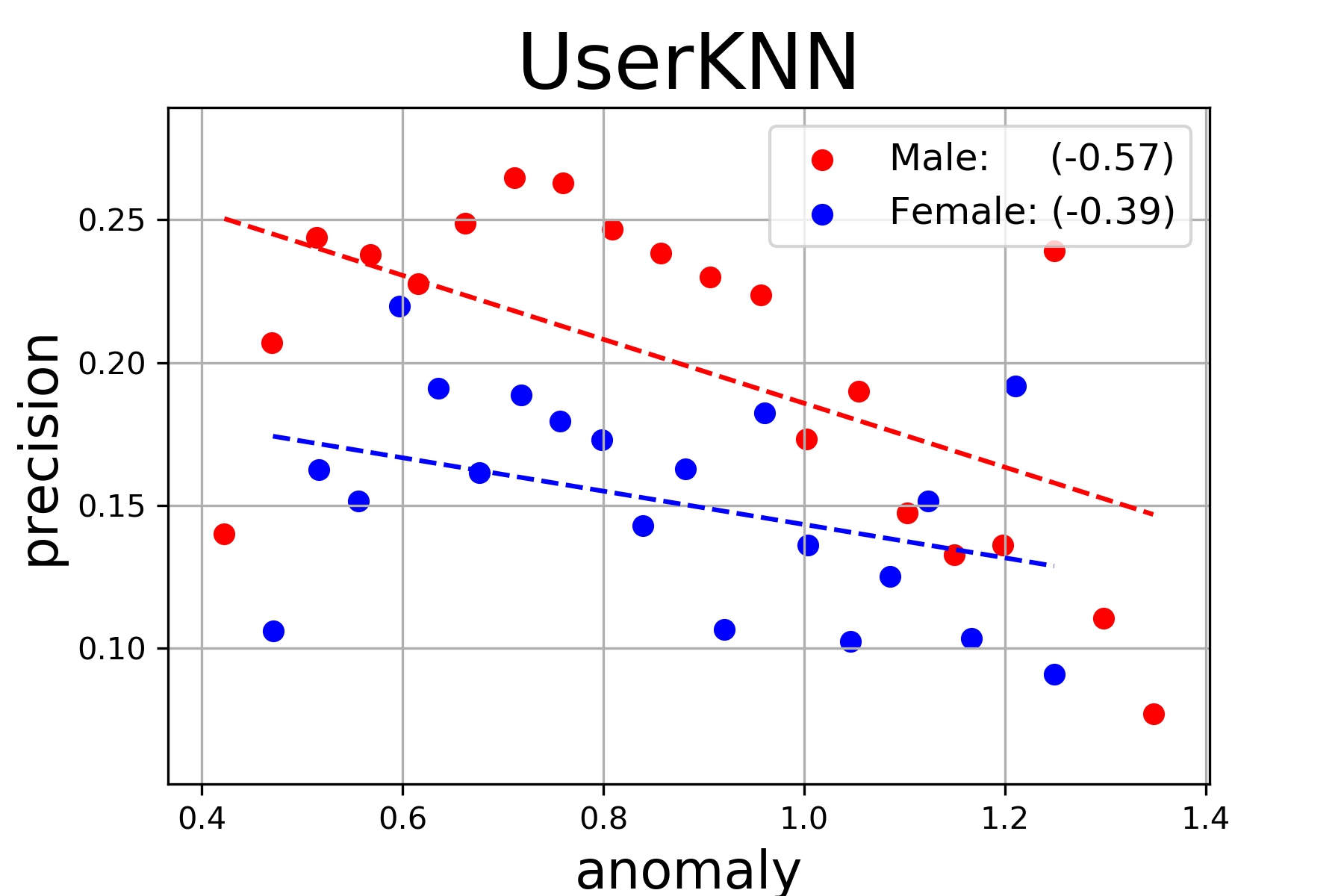}
        \includegraphics[width=0.24\textwidth]{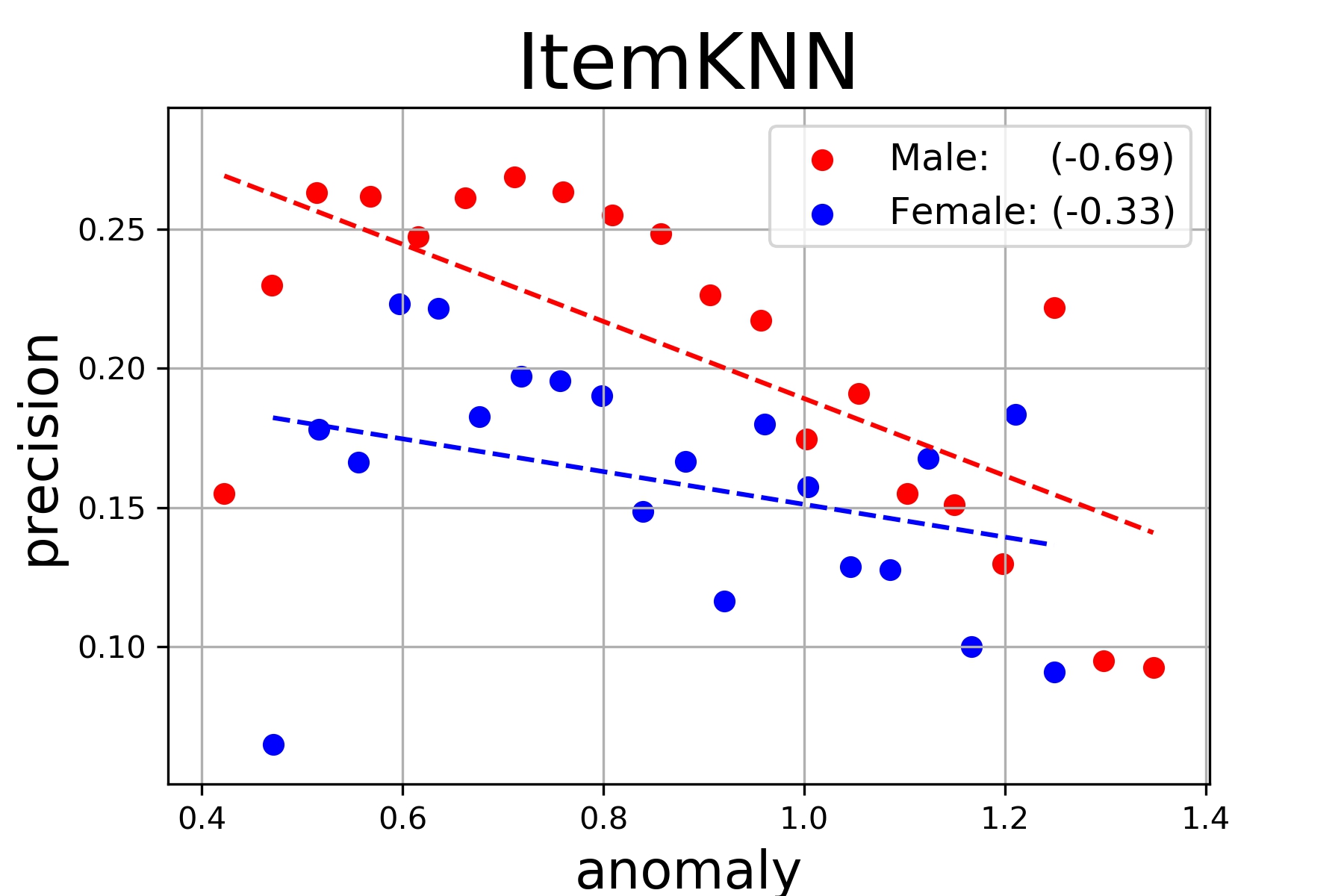}
        \includegraphics[width=0.24\textwidth]{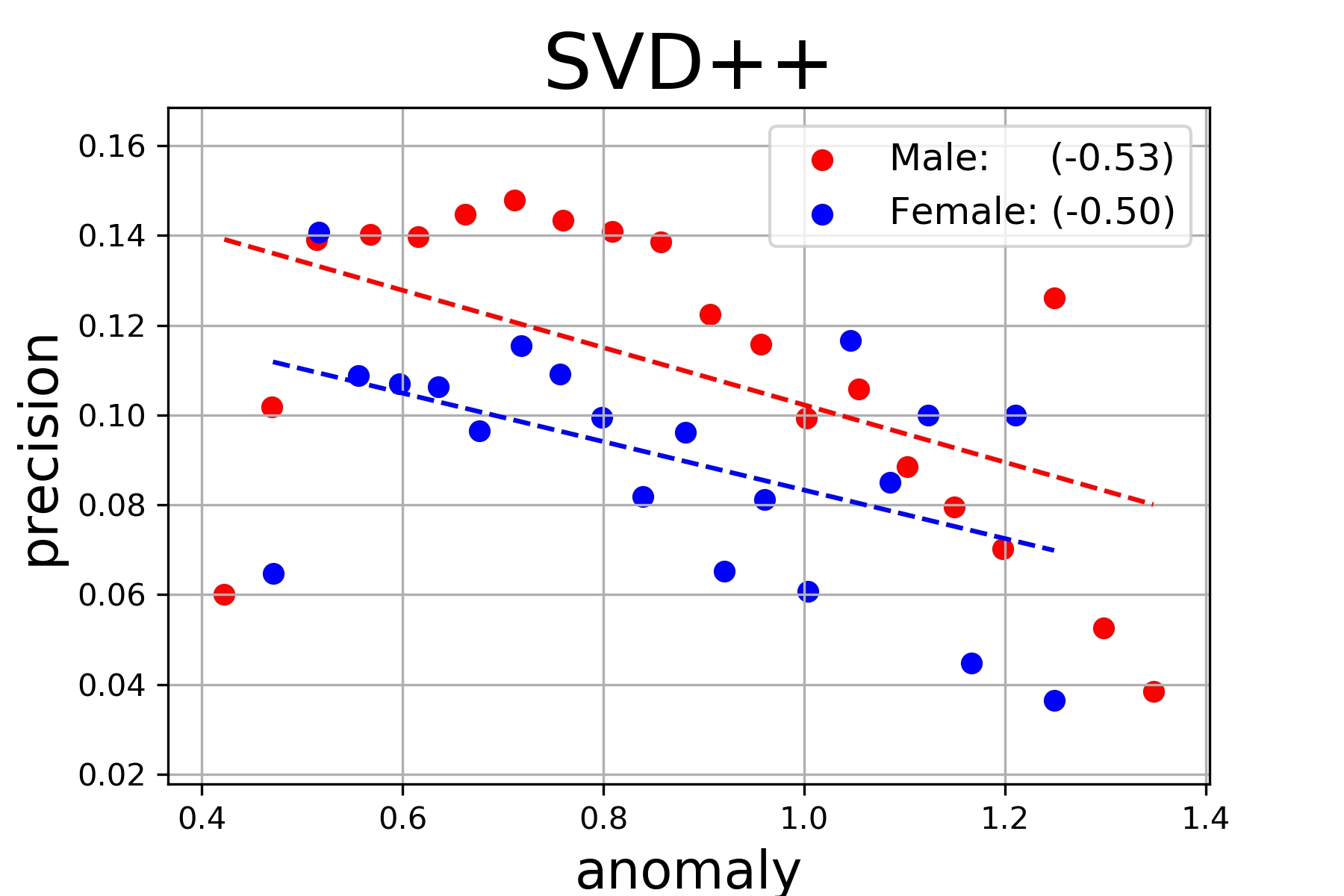}
        \includegraphics[width=0.24\textwidth]{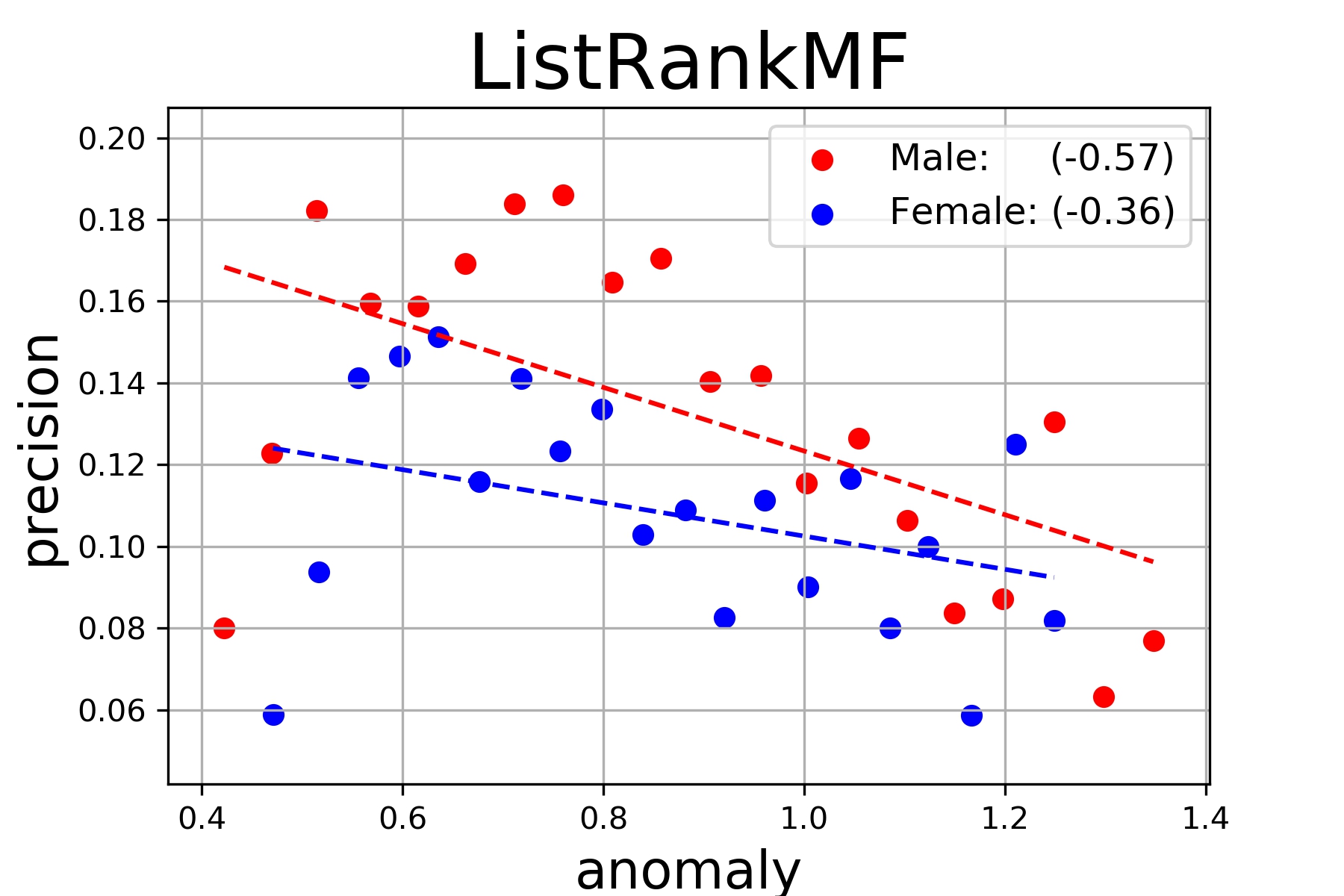}
        
  \end{subfigure}
  \begin{subfigure}[b]{0.99\textwidth}
        \includegraphics[width=0.24\textwidth]{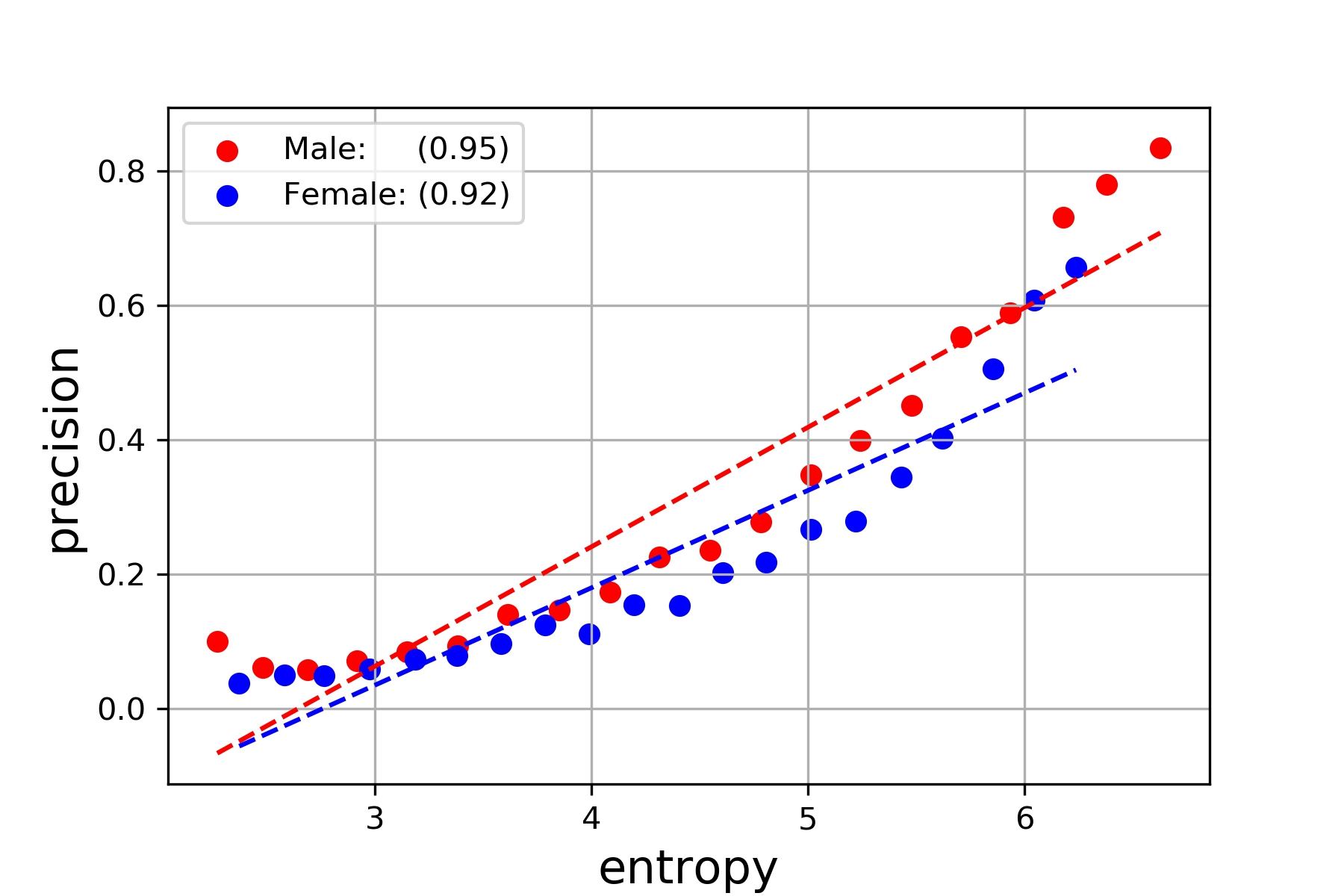}
        \includegraphics[width=0.24\textwidth]{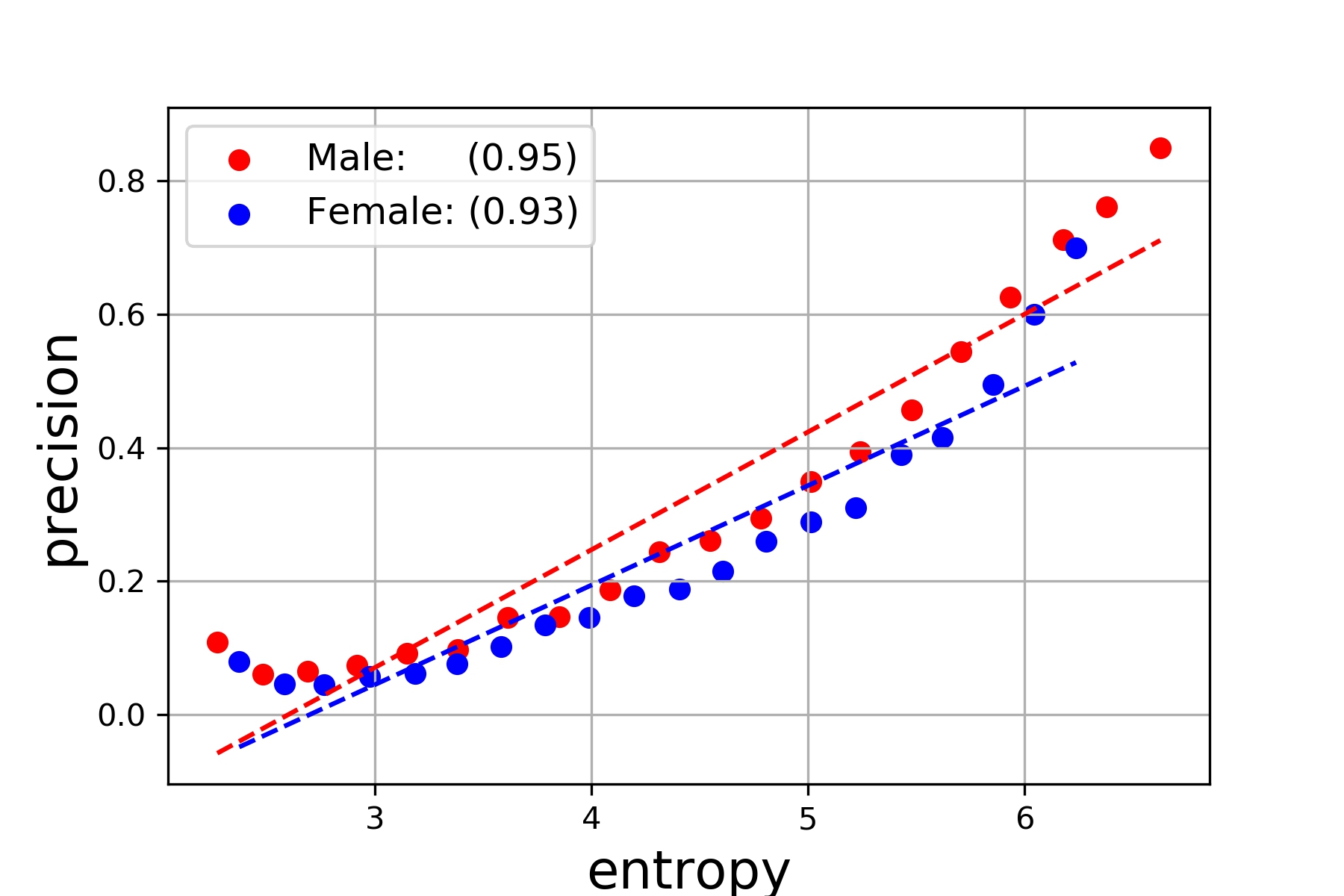}
        \includegraphics[width=0.24\textwidth]{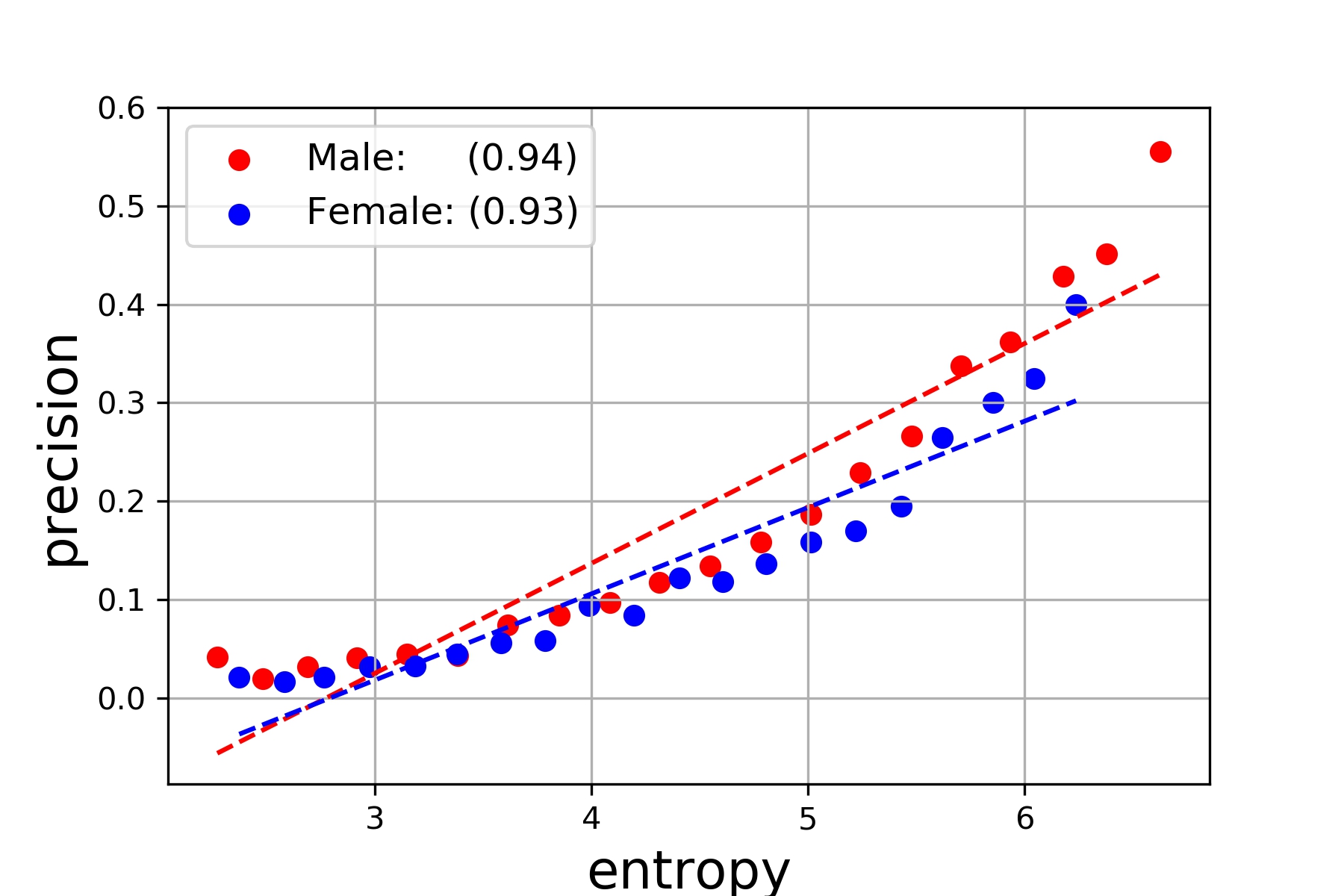}
        \includegraphics[width=0.24\textwidth]{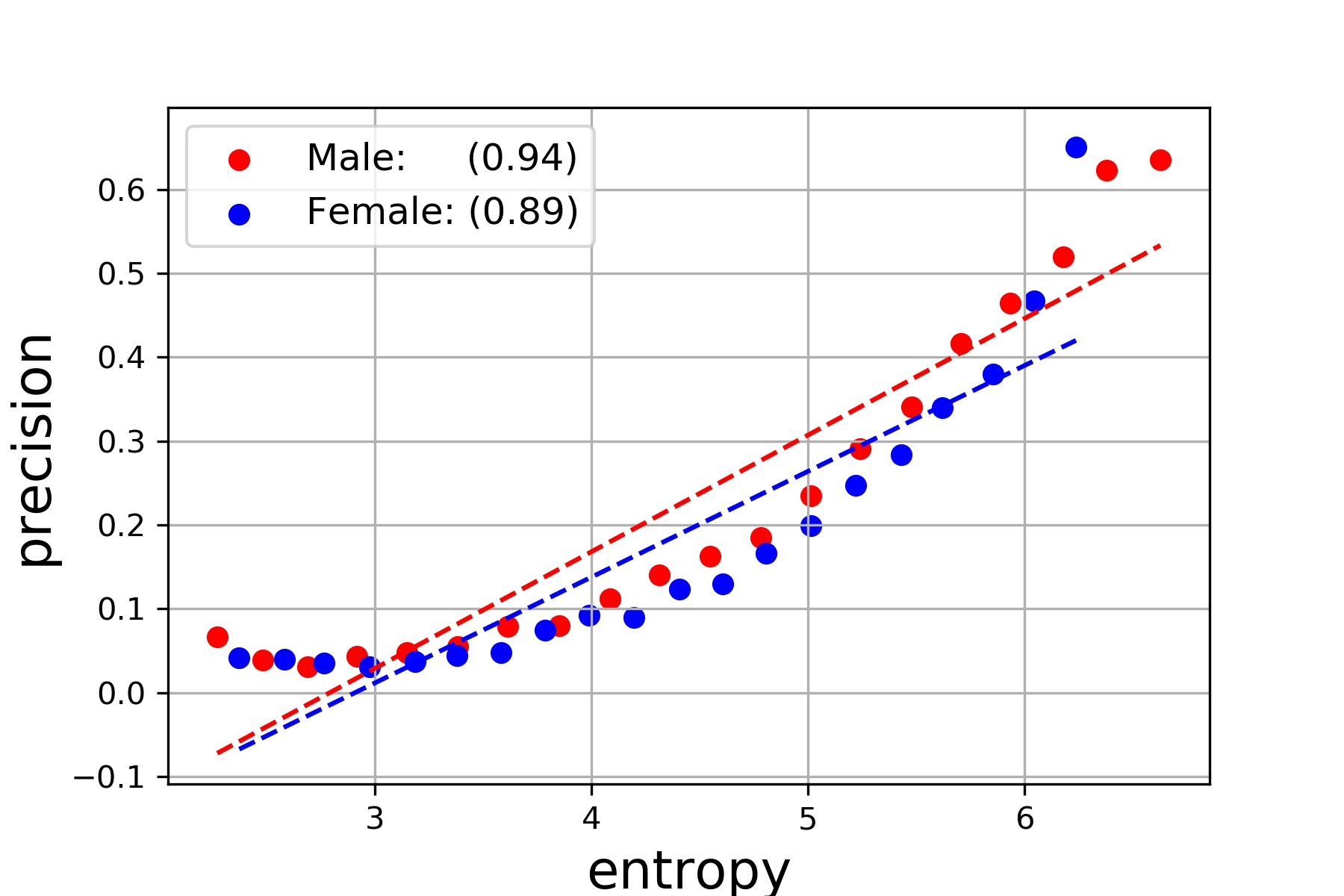}
  \end{subfigure}
  \begin{subfigure}[b]{1\textwidth}
        \includegraphics[width=0.24\textwidth]{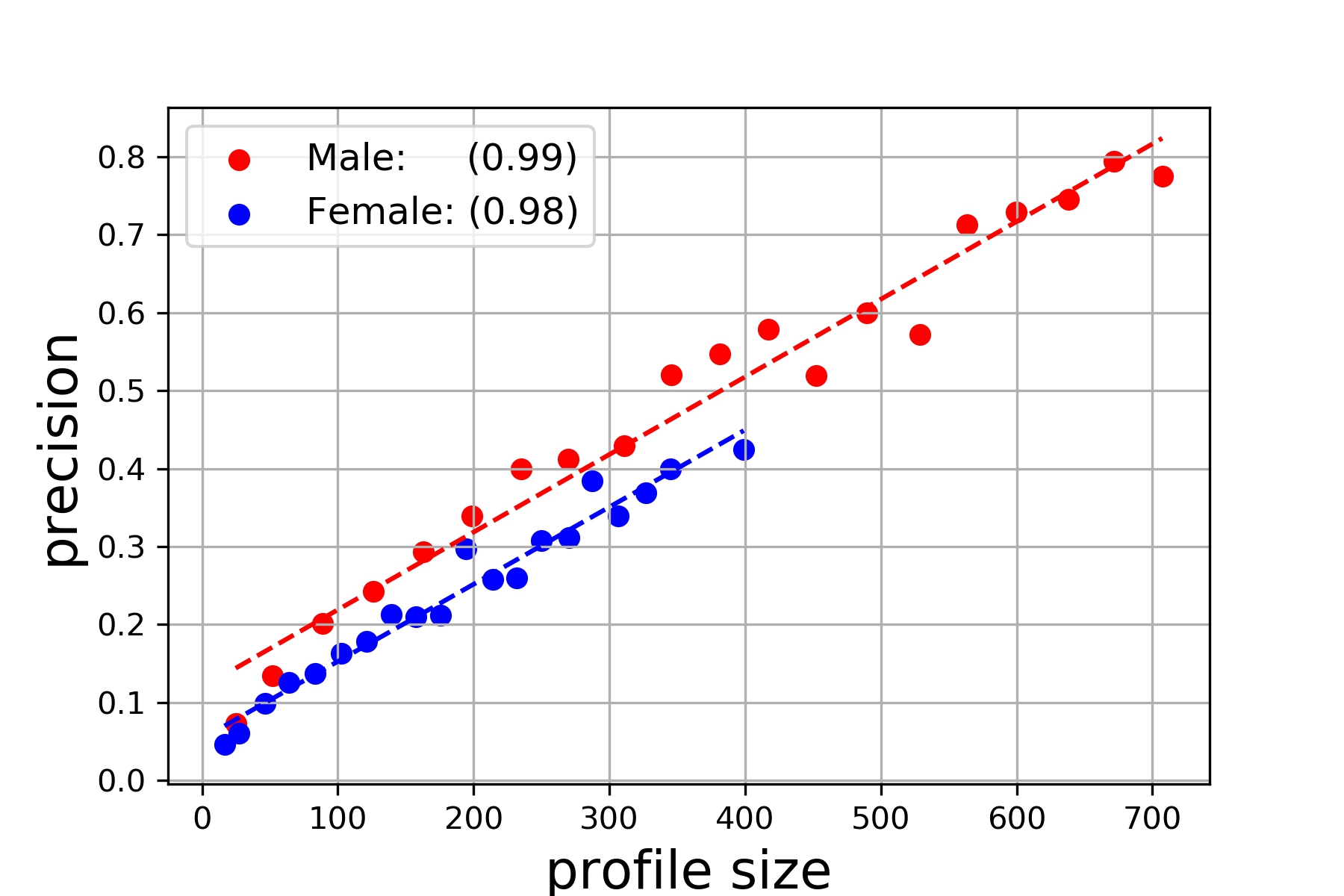}
        \includegraphics[width=0.24\textwidth]{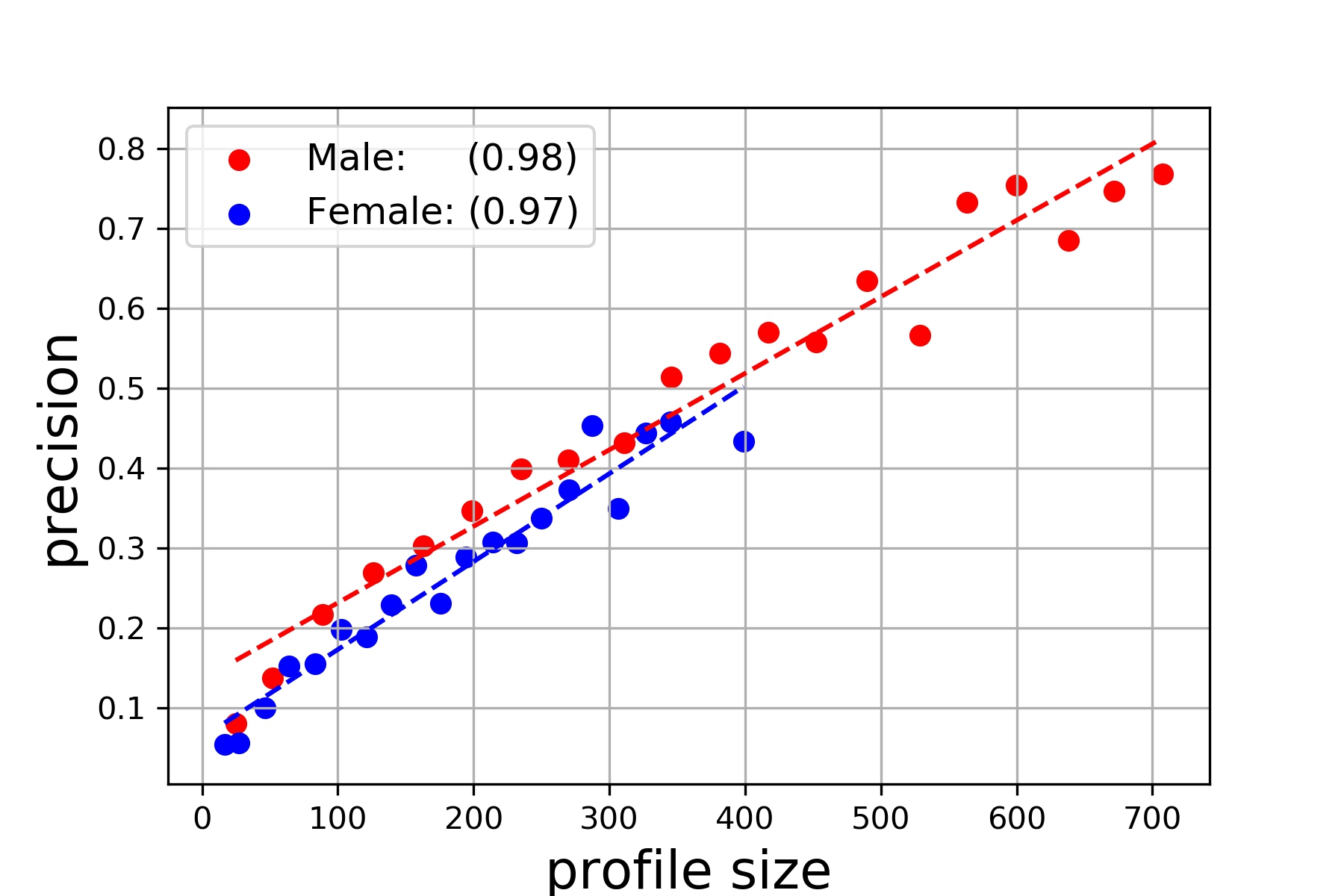}
        \includegraphics[width=0.24\textwidth]{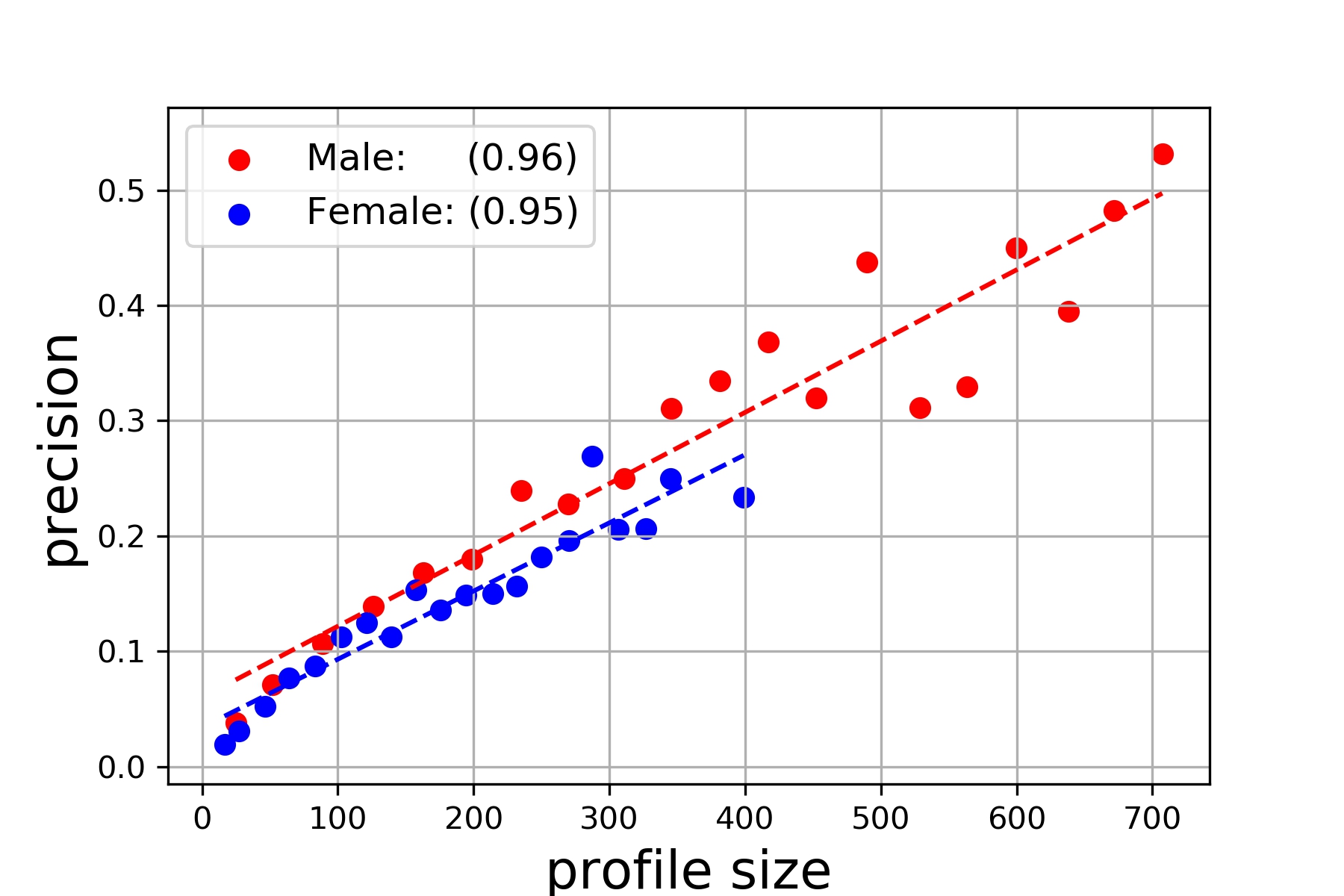}
        \includegraphics[width=0.24\textwidth]{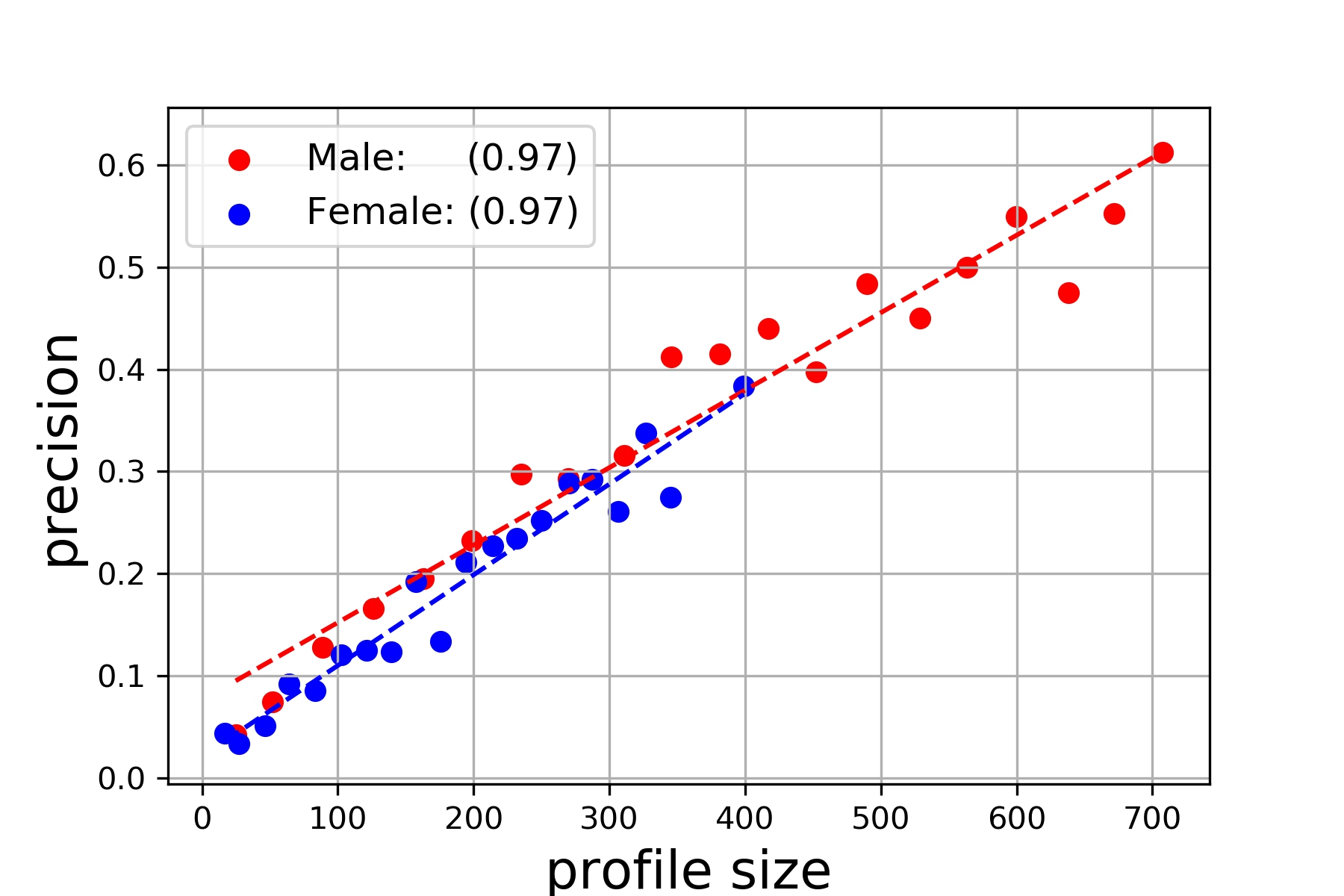}
  \end{subfigure}
\caption{The Correlation between anomaly, entropy, and size of the users' profiles and precision of the recommendations generated for them. Numbers next to the legends in the plots show the correlation coefficient for each user group.} \label{fig:corr_precs}
\end{figure*}

Figure~\ref{fig:corr_precs} shows the correlation between the aforementioned factors for different user groups and the precision of their recommendations. Unlike miscalibration, it seems the correlations of these three factors with precision are much stronger. For example, the first row of this Figure shows that the higher the inconsistency of the ratings, the lower the precision is, which is what we expected. The second row shows a strong correlation (correlation coefficient $\approx$ 0.9) indicating that user groups with higher entropy (more information gain) in their ratings receive more accurate (higher precision) recommendations. Also, from the same Figure, we can see for the lower values of entropy, the algorithms behave more fairly, but, the larger the entropy gets, the discrimination between female and male user groups becomes more apparent (higher precision for male user groups). The relationship between average profile size and precision is also shown in the last row of Figure~\ref{fig:corr_precs}. As expected, user groups with larger profiles benefit from more accurate recommendations for both males and females. However, the discrimination can still be seen for some algorithms such as \algname{UserKNN} where female users with the same profile size still receive recommendations with lower precision compared to the male users. 

\section{Conclusion and Future Work}
Unfairness in recommendation is an important issue that needs to be diagnosed and treated properly. In this paper, we investigated the impact of three different factors on recommendation performance and fairness: the degree of anomaly, entropy, and profile size. We made several interesting observations that need further research. 
First, we observed that neighborhood-based algorithms such as \algname{UserKNN} and \algname{ItemKNN} discriminate more against women in this data set, in part because they are more affected by the specific characteristics of the profiles such as profiles size and entropy as these characteristics are present more in the female profiles than male ones. On the other hand, the Matrix Factorization based methods (\algname{ListRankMF} and \algname{SVD++}) that rely on lower dimensional latent space and focus more on user-item interactions seem to even out this effect and make the recommendations fairer across different genders. In particular, the \algname{SVD++} algorithm was able to give a comparable performance for both male and female users, which provided a more \textit{fair} treatment across genders than the other algorithms.
For future work, we intend to use more datasets for our analysis. We will also investigate the contribution of each of the mentioned factors on recommendation performance. Finally, we will explore other potential factors that could play a role in the unfairness of recommendation algorithms. 

\bibliographystyle{aaai}
\bibliography{main}

\begin{thebibliography}{}

\bibitem[\protect\citeauthoryear{Abdollahpouri \bgroup et al\mbox.\egroup
  }{2019a}]{abdollahpouriWSDM}
Abdollahpouri, H.; Mansoury, M.; Burke, R.; and Mobasher, B.
\newblock 2019a.
\newblock The impact of popularity bias on fairness and calibration in
  recommendation.
\newblock {\em arXiv preprint arXiv:1910.05755}.

\bibitem[\protect\citeauthoryear{Abdollahpouri \bgroup et al\mbox.\egroup
  }{2019b}]{abdollahpourirmse1}
Abdollahpouri, H.; Mansoury, M.; Burke, R.; and Mobasher, B.
\newblock 2019b.
\newblock The unfairness of popularity bias in recommendation.
\newblock {\em In Workshop on Recommendation in Multistakeholder Environments
  (RMSE)}.

\bibitem[\protect\citeauthoryear{Ekstrand \bgroup et al\mbox.\egroup
  }{2018}]{ekstrand2018}
Ekstrand, M.~D.; Tian, M.; Azpiazu, I.~M.; Ekstrand, J.~D.; Anuyah, O.;
  McNeill, D.; and Pera, M.~S.
\newblock 2018.
\newblock All the cool kids, how do they fit in?: Popularity and demographic
  biases in recommender evaluation and effectiveness.
\newblock In {\em In Conference on Fairness, Accountability and Transparency},
  172--186.

\bibitem[\protect\citeauthoryear{Guo \bgroup et al\mbox.\egroup
  }{2015}]{Guo2015}
Guo, G.; Zhang, J.; Sun, Z.; and Yorke-Smith, N.
\newblock 2015.
\newblock Librec: A java library for recommender systems.
\newblock In {\em UMAP Workshops}.

\bibitem[\protect\citeauthoryear{Kamishima, Akaho, and
  Sakuma}{2011}]{kamishima2011}
Kamishima, T.; Akaho, S.; and Sakuma, J.
\newblock 2011.
\newblock Fairness-aware learning through regularization approach.
\newblock In {\em In 11th International Conference on Data Mining Workshops},
  643--650.

\bibitem[\protect\citeauthoryear{Mansoury \bgroup et al\mbox.\egroup
  }{2018}]{mansoury2018automating}
Mansoury, M.; Burke, R.; Ordonez-Gauger, A.; and Sepulveda, X.
\newblock 2018.
\newblock Automating recommender systems experimentation with librec-auto.
\newblock In {\em Proceedings of the 12th ACM Conference on Recommender
  Systems},  500--501.
\newblock ACM.

\bibitem[\protect\citeauthoryear{Mansoury \bgroup et al\mbox.\egroup
  }{2019}]{mansoury2019relationship}
Mansoury, M.; Abdollahpouri, H.; Rombouts, J.; and Pechenizkiy, M.
\newblock 2019.
\newblock The relationship between the consistency of users' ratings and
  recommendation calibration.
\newblock {\em arXiv preprint arXiv:1911.00852}.

\bibitem[\protect\citeauthoryear{Yao and Huang}{2017}]{yao2017}
Yao, S., and Huang, B.
\newblock 2017.
\newblock Beyond parity: Fairness objectives for collaborative filtering.
\newblock In {\em In Advances in Neural Information Processing Systems},
  2921--2930.

\bibitem[\protect\citeauthoryear{Zhu, Hu, and Caverlee}{2018}]{zhu2018}
Zhu, Z.; Hu, X.; and Caverlee, J.
\newblock 2018.
\newblock Fairness-aware tensor-based recommendation.
\newblock In {\em In Proceedings of the 27th ACM International Conference on
  Information and Knowledge Management},  1153--1162.

\end{thebibliography}
\end{document}